\documentclass[aps,pra,amsmath,amssymb,reprint,showpacs]{revtex4-1}

\usepackage{graphicx}% Include figure files
\usepackage{dcolumn}% Align table columns on decimal point
\usepackage{bm}% bold math

\def\ket#1{\mathinner{|{#1}\rangle}}

\begin{document}

\title{Polarization-rotation resonances with subnatural widths using a control laser}

 \author{Sapam Ranjita Chanu}
 \affiliation{Department of Physics, Indian Institute of
 Science, Bangalore 560\,012, India}
 \author{Kanhaiya Pandey}
 \altaffiliation{Centre for Quantum Technologies, National University of Singapore, Singapore}
 \author{Vineet Bharti}
 \altaffiliation{Department of Physics, Indian Institute of Technology, Roorkee 247\,667, India}
 \author{Ajay Wasan}
 \altaffiliation{Department of Physics, Indian Institute of Technology, Roorkee 247\,667, India}
 \author{Vasant Natarajan}
 \affiliation{Department of Physics, Indian Institute of
 Science, Bangalore 560\,012, India}
 \email{vasant@physics.iisc.ernet.in}
 \homepage{www.physics.iisc.ernet.in/~vasant}

\begin{abstract}
We demonstrate extremely narrow resonances for polarization rotation in an atomic vapor. The resonances are created using a strong control laser on the same transition, which polarizes the atoms due to optical pumping among the magnetic sublevels. As the power in the control laser is increased, successively higher-order nested polarization rotation resonances are created, with progressively narrower linewidths. We study these resonances in the $D_2$ line of Rb in a room temperature vapor cell, and demonstrate a width of $0.14 \, \Gamma$ for the third-order rotation.  {The physical basis for the observed resonances is that optical pumping results in a simplified $\Lambda$V-type level structure {\em with differential dressing of the levels by the control laser}, which is why the control power has to be sufficiently high for each resonance to appear. This explanation is borne out by a density-matrix analysis of the system.} The dispersive lineshape and subnatural width of the resonance lends itself naturally to applications such as laser locking to atomic transitions and precision measurements.
\end{abstract}

\pacs{32.80.Qk, 33.55.+b, 32.80.Xx}

%\pacs{32.80.Qk}{Coherent control of atomic interactions with photons}
%\pacs{33.55.+b}{Optical activity and dichroism}
%\pacs{32.80.Xx}{Level crossing and optical pumping}

\maketitle

\section{Introduction}

Rotation of the plane of polarization of linearly polarized light as it propagates through a medium, known as {\em optical rotation}, is a well-known phenomenon that occurs because the medium has different refractive indices for the two opposite components of circular polarization. This difference (or birefringence) can be created in a variety of ways---for example, by use of a magnetic field in the phenomenon of magneto-optic rotation (MOR), which is called the Faraday effect when the applied field is longitudinal and the Voigt effect when the applied field is transverse.

Optical fields can also be used to induce anisotropy in an atomic medium---through optical pumping by circularly polarized light \cite{CDU72} or by elliptically polarized light \cite{RHB01}, or through resonant two-photon dispersion \cite{LIB76}, {\em etc}. In ``two-level'' systems (but with multiple magnetic sublevels due to degeneracy), the use of a counter-propagating circularly polarized pump beam has been used to observe Doppler-free rotation resonances in what is called saturated polarization spectroscopy \cite{WIH76}. {This is similar to the standard technique of saturated absorption spectroscopy, and gives linewidths that are 2 to 3 times the natural linewidth $\Gamma$ \cite{BAN03}. The only resonances with subnatural linewidths in previous studies have been observed in a similar polarization spectroscopy experiment, but with a strong probe beam \cite{GKT82}}. In three-level systems, a control laser on an auxiliary transition has been used to control the state of polarization of a probe beam on a primary transition, in ladder-type \cite{WIG98,PBC97}, $\Lambda$-type \cite{WLM06}, and {inverted Y-type \cite{DPG09}} systems. In earlier work from our laboratory, we studied rotation in the presence of both a control laser and a magnetic field \cite{PWN08}, which we call coherent control of MOR. We found an interesting interplay between the two effects with regions of suppressed and enhanced rotation.

In this work, we study optical rotation of a linearly polarized probe laser passing through atomic vapor, and observe resonances with widths significantly narrower than the natural linewidth. The rotation is induced by a strong co-propagating circularly polarized control laser driving the same transition, similar to the phenomenon of electromagnetically induced transparency (EIT) \cite{FIM05} and electromagnetically induced absorption (EIA) \cite{CSB11}. In both cases, the control laser modifies the susceptibility of the medium. But in EIT and EIA, it is the imaginary part of the susceptibility that plays a role because we are considering the {\em absorption} of the probe laser, while in the experiments here it is the real part of the susceptibility that is relevant because we are looking at the {\em phase change} of the probe laser.

The normal control-induced optical rotation, as has been used in earlier experiments \cite{WIG98,PWN08}, is caused because the control laser polarizes the atoms. This is due to optical pumping among the ground-state magnetic sublevels, so that all the population is in the extreme (stretch) sublevel. \emph{However, this rotation is generally not subnatural}. As the power in the control laser is increased,  {the sublevels get ``dressed'' by the control laser. It is the dressing that results in the appearance of the subnatural resonances, which appear nested within the primary rotation spectrum.}

We study these resonances in the $D_2$ line of $^{87}$Rb in a  room temperature vapor cell, using the $F=2 \rightarrow F'=3$ hyperfine transition. The relevant sublevels after optical pumping form a $\Lambda$V-type system. But the dressed-state separation for the two sets of levels coupled by the control laser are different because of the different relative strengths of the two transitions.  {This differential dressing of the levels is the underlying reason for the appearance of the higher-order rotation signals, an explanation that is borne out by a density-matrix calculation of the rotation in a $\Lambda$V-type system, with and without the $\Lambda$ part.} Both the second-order and third-order rotations are subnatural, with widths as small as $0.9 \, \Gamma$ and $0.14 \, \Gamma$ respectively.

\section{Experimental details}

\begin{figure}
\centering{\resizebox{0.95\columnwidth}{!}{\includegraphics{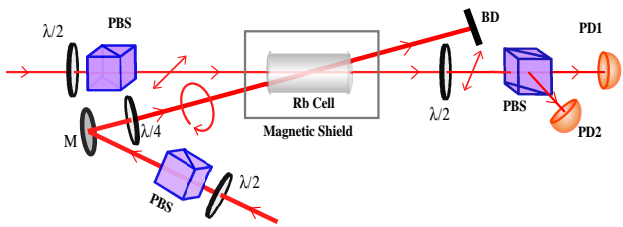}}}
\caption{(Color online) Schematic of the experiment. The control laser is circularly polarized. The probe laser is linearly polarized, and its two orthogonal linear components are separately detected. The angle between the beams is exaggerated for clarity, in reality it is only about 10 $\mu$rad. Figure key: $\lambda/2$ -- halfwave retardation plate, $\lambda/4$ -- quaterwave retardation plate, PBS -- polarizing beamsplitter cube, M -- mirror, BD -- beam dump, PD -- photodiode.}
 \label{setup}
\end{figure}

A schematic of the experiment is shown in Fig.\ \ref{setup}. The probe and control beams are derived from two {\em independent} grating-stabilized diode lasers operating on the $D_2$ line of Rb at 780 nm. The output beam of both lasers is elliptic, with $1/e^2$ size of 3 mm $\times$ 4 mm. The probe beam is apertured to a circular size of 2 mm so that it overlaps near the center of the control beam. The power in each beam is controlled independently by using a halfwave retardation plate in front of a polarizing beamsplitter cube (PBS). The probe beam is locked to the $F=2 \rightarrow F'=3$ hyperfine transition of $^{87}$Rb using modulation-free polarization spectroscopy in a vapor cell. The control laser is scanned around the same transition. {This technique of locking the probe while scanning the control is a standard technique that we use in these studies \cite{CSB11,CPN12}, which has the main advantage of making the probe spectrum appear on a flat Doppler-free background \cite{BAN03}, but does not affect the results in any other way.}

The probe beam is linearly polarized, while the control beam is right circularly polarized. The two beams co-propagate through a cylindrical room temperature Rb vapor cell of dimensions 2.5 cm $\times$ 5 cm, with a small angle of about 10 $\mu$rad. The cell has pure elemental Rb with no buffer gas and no anti-relaxation coating on the wall. It is placed inside a magnetic shield, which reduces the ambient field to less than 0.5 mG. The probe beam is separated into its horizontal and vertical components using a PBS, and the two components are detected using photodiodes. The halfwave plate before the PBS is used to balance the two photodiode signals (by making the polarization roughly at $45^\circ$) {\em in the absence of the control beam but with all the other optical elements in place}, so as to account for residual rotation due to any birefringence at the cell walls and other surfaces. The rotation angle is calculated from the two photodiode intensities as:
\begin{equation}
 \label{rotangle}
\phi=\frac{1}{2} \arcsin {\frac{I_{1}-I_{2}}{I_{1}+I_{2}}}
\, ,
\end{equation}

\section{Experimental Results}

\begin{figure}
\centering{\resizebox{0.95\columnwidth}{!}{\includegraphics{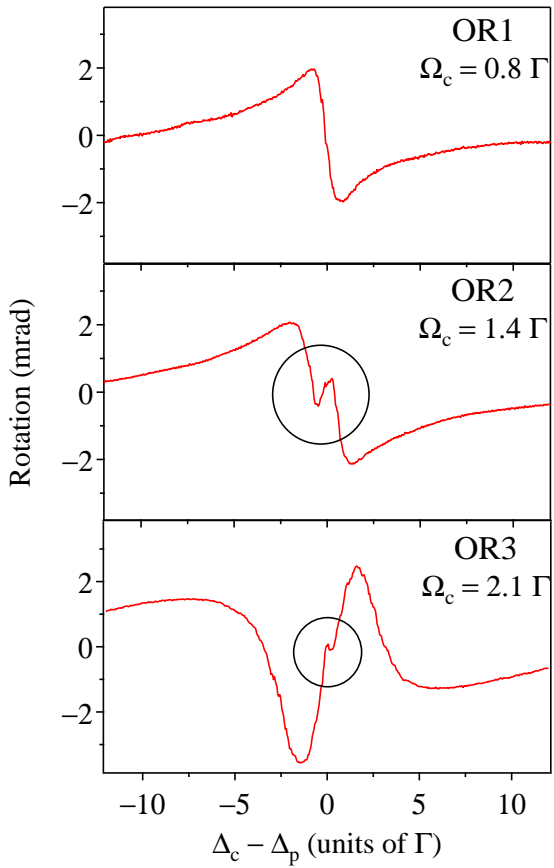}}}
\caption{(Color online) Polarization-rotation spectra vs.\ relative detuning between the two lasers, for different values of control power (Rabi frequency). The top panel shows the normal rotation (OR1) seen at low power. The middle panel is the second-order rotation (OR2) obtained at a slightly higher power. The bottom panel shows the third-order rotation (OR3) signal when the control power is increased further. Each successive signal is nested within the previous ones.}
 \label{rot}
\end{figure}

A typical rotation spectrum in the presence of the control beam is shown in Fig.\ \ref{rot}. The scan axis is shown as the relative detuning between the probe and control lasers, though in the experiment only $\Delta_c$ is varied while $\Delta_p$ is fixed at 0. There are three regions, as shown in the three panels of the figure. First, at a low control power (corresponding to a maximum Rabi frequency of $0.8 \, \Gamma$ at the center of the beam), we get the first-order optical rotation (OR1), which, as explained before, is due to the different control-induced susceptibilities for the $\sigma^+$ and $\sigma^-$ components of the probe laser. Second, at a slightly higher power of $1.4 \, \Gamma$, a narrower second-order optical rotation signal (OR2) nested within the previous one begins to appear. The OR2 signal has an opposite sign from the OR1 signal. The third-order rotation signal (OR3) appears when the power is increased further beyond $2.1 \, \Gamma$. The signal (OR3) is narrower still, and is nested within the other two signals. It has the opposite sign compared to the OR2 signal, and the same sign as the OR1 signal. Thus, each successive rotation has the opposite sign compared to the previous one. At the power where the OR3 signal starts to appear, the OR2 signal becomes quite large and dominates over the OR1 signal.

\begin{figure}
\centering{\resizebox{0.9\columnwidth}{!}{\includegraphics{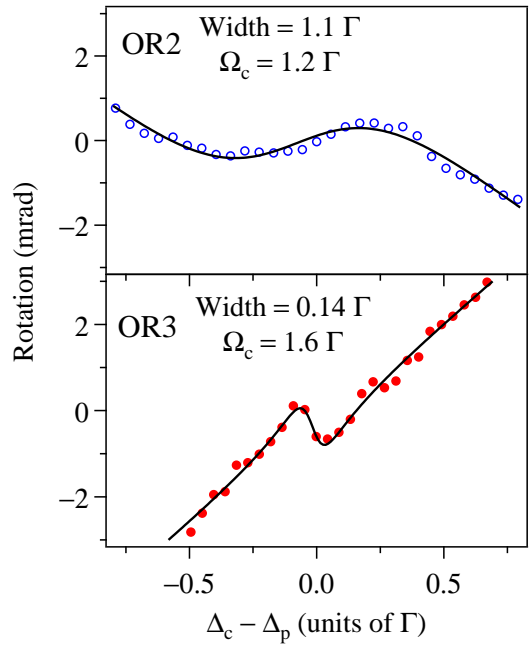}}}
\caption{(Color online) Zoomed-in versions of the higher-order rotation signals seen in Fig.\ \ref{rot}. The circles represent the measured spectra, and the solid lines are curve fits to a dispersive lineshape along with a linear part to account for the previous rotation signal on which it is nested, as described in the text. The maximum control Rabi frequency at beam center, and the linewidth from the curve fit are listed.}
 \label{zoom}
\end{figure}

Zoomed-in versions of the higher-order rotations are shown in Fig.\ \ref{zoom}. The spectra are taken at a slightly lower power compared to that used for the earlier figure. The circles represent the measured spectrum, while the solid line is a curve fit to the following equation
\begin{equation}
\phi = \frac{Ax}{x^2 + (\Gamma/2)^2} + Bx \, ,
 \label{fit}
\end{equation}
i.e.\ a dispersive lineshape of width $\Gamma$, along with a linear part to account for the previous rotation within which each signal is nested.
Both the OR2 and OR3 signals are well described by this lineshape. More interestingly, the OR3 signal has a subnatural width of only $0.14 \, \Gamma$ (0.82 MHz), which is 2 to 3 times smaller than the linewidth seen in EIT and EIA experiments on the same system \cite{CSB11}. In addition, the dispersive lineshape lends itself naturally to applications such as tight locking of lasers to an atomic transition {\em without modulation}.

\begin{figure}
\centering{\resizebox{0.95\columnwidth}{!}{\includegraphics{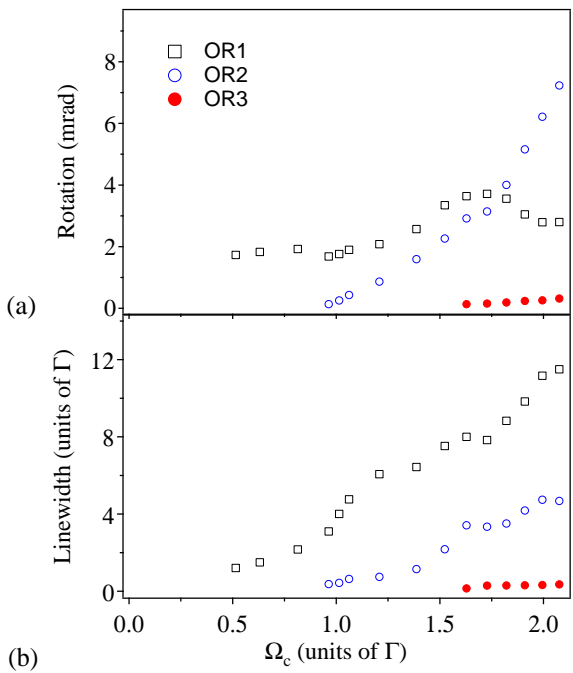}}}
\caption{{(Color online) Amplitude and linewidths for the three kinds of rotation, plotted as a function of control Rabi frequency at beam center. (a) The amplitude of the OR2 signal starts to dominate over the OR1 signal from the point where the OR3 signal first appears. (b) The OR2 linewidth is subnatural for some part, while the OR3 linewidth remains subnatural over the entire range.}}
 \label{widthamp}
\end{figure}

{The amplitude and linewidths obtained from the curve fits to the three different rotation signals are shown in Fig.\ \ref{widthamp}. They are plotted as a function of the maximum control Rabi frequency at beam center. As explained before, the control power has to be high enough for the different rotation signals to appear. As a result, the OR2 signal begins to appear only after $\Omega_c = 0.9 \, \Gamma$, while the OR3 signal begins to appear after a still higher power of $\Omega_c = 1.6 \, \Gamma$. The width of the OR2 resonance is subnatural when it first appears, but becomes larger than $\Gamma$ after $\Omega_c = 1.2 \, \Gamma$. On the other hand, the OR3 resonance remains subnatural over the entire range studied, demonstrating its power for precision measurements. Interestingly, from the point at which the OR3 signal begins to appear, the amplitudes of the OR1 and OR2 signals become roughly equal, and the OR2 signal soon begins to dominate.}

\section{Theoretical analysis}

\begin{figure}
\centering{\resizebox{0.95\columnwidth}{!}{\includegraphics{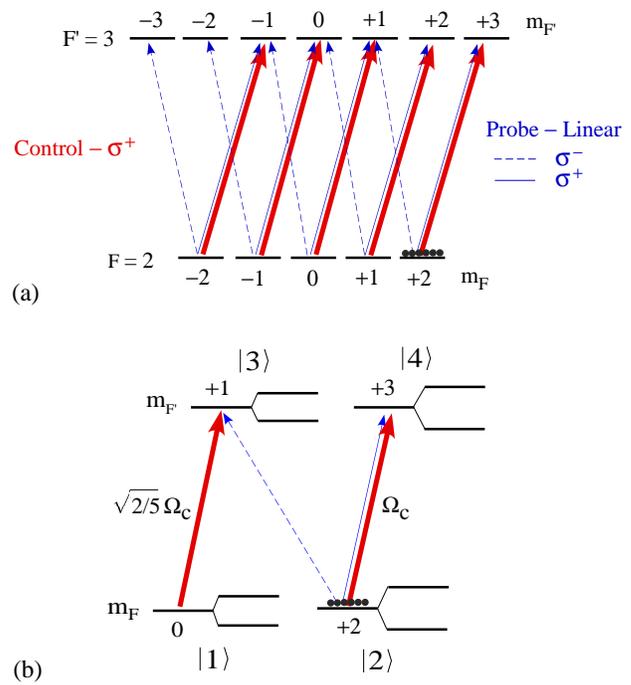}}}
\caption{(Color online) (a) Complete magnetic sublevels of the $F=2 \rightarrow F'=3$ hyperfine transition in $^{87}$Rb. The sublevels coupled by the right circularly polarized control beam, and the two circular components of the linearly polarized probe beam, are shown. (b) Due to optical pumping by the control laser, all the population is in the extreme sublevel, so that the level structure {becomes the simplified $\Lambda$V-type system shown here.} The dressed-state separation is different because of the different Rabi frequencies as determined by the respective Clebsch-Gordan coefficients. }
 \label{levels}
\end{figure}

In order to understand the observed rotation signal theoretically, we have done a density-matrix analysis of the sublevel structure. Due to optical pumping by the circularly polarized control laser, the entire population is in the extreme $\ket{F=2,m_F=2}$ sublevel. {Therefore, the level structure is simplified to the $\Lambda$V-type system, as shown in Fig.\ \ref{levels}}. The relevant sublevels in the simplified system are labeled $\ket{1}$ to $\ket{4}$, so that the control laser couples the $\ket{1} \rightarrow \ket{3}$ and $\ket{2} \rightarrow \ket{4}$ transitions, while the probe laser couples the $\ket{2} \rightarrow \ket{3}$ transition for its left-circular component and $\ket{2} \rightarrow \ket{4}$ transition for its right-circular component. It is important to note that the dressed-state separation (equal to the AC Stark shift) is different for the two transitions because they have different Rabi frequencies, as determined by the respective Clebsch-Gordan coefficients. {Thus, the $\ket{1} \rightarrow \ket{3}$ transition has a Rabi frequency of $\sqrt{2/5} \, \Omega_c$, while the $\ket{2} \rightarrow \ket{4}$ transition has a Rabi frequency of $\Omega_c$}. This is shown schematically (not to scale) in Fig.\ \ref{levels}(b).

The decay rate of the excited levels $\ket{3}$ and $\ket{4}$ is taken to be $2 \pi \times 6$ MHz, corresponding to the linewidth of the $5P_{3/2}$ state. Decayed atoms from these levels are added to the populations in the ground levels $\ket{1}$ and $\ket{2}$ with the corresponding branching ratios, {i.e.\ one third of the atoms decaying from $\ket{3}$ are added to both populations, while all atoms decaying from $\ket{4}$ are added to the population of $\ket{2}$}. The relaxation rate of coherence between these ground levels is taken to be $2 \pi \times 0.1$ MHz, in order to account for the finite linewidth of the control laser and transit-time broadening \cite{BMW09,LCW12}. The various density-matrix elements in steady state and using the rotating wave approximation are solved {\em numerically} following the procedure given by one of the authors in Ref.\ \cite{PAN13}. This procedure is similar to calculations done before in $N$-type systems \cite{GWR04} and doubly driven V-type systems \cite{EGD01}.

\begin{figure}
\centering{\resizebox{0.8\columnwidth}{!}{\includegraphics{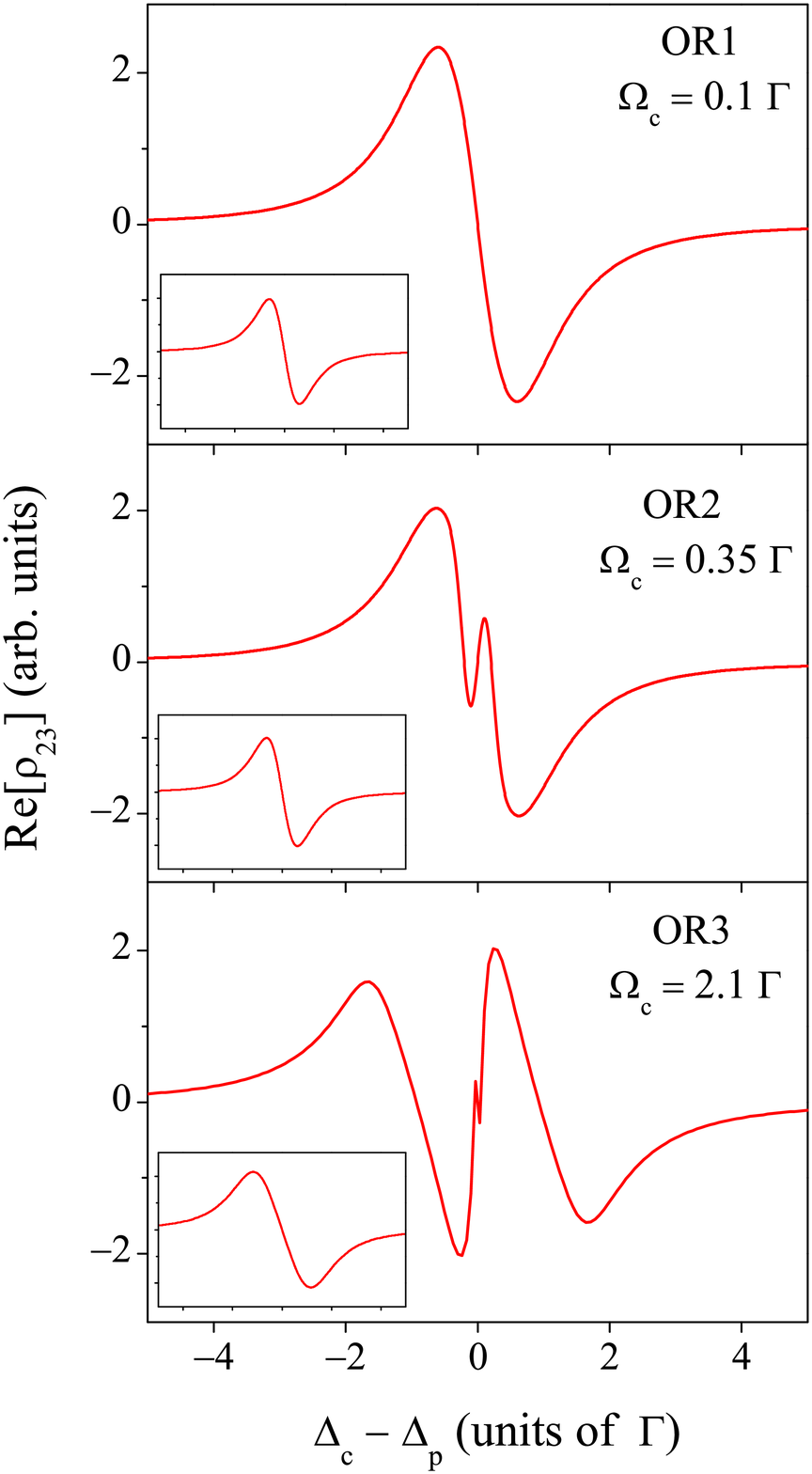}}}
\caption{(Color online) Calculated probe rotation (Re$[\rho_{23}]$) for three values of the control Rabi frequency. {The insets show the results of the calculation on the same scale when the $\Lambda$-type system is ignored, and only the V-type is included.}}
 \label{theor}
\end{figure}

Since the rotation is due to the difference between the real part of the susceptibilities for the $\sigma^+$ and $\sigma^-$ components of the probe, it is proportional to Re[$\rho_{24}-\rho_{23}$].
With the strong control laser on the $\ket{2} \rightarrow \ket{4}$ transition, the weak probe laser has negligible effect on $\rho_{24}$. Hence, the measured rotation signal is entirely due to Re$[\rho_{23}]$. The calculated value of this quantity for three values of control Rabi frequency, {\em after taking Doppler averaging into account}, is shown in Fig.\ \ref{theor}.

Consistent with our experimental results, the calculated spectra reproduce the appearance of three rotations as the control power is increased. Each higher-order rotation is nested within the previous one and has the opposite sign, exactly as observed experimentally. In addition, the calculation reproduces the observation that the amplitude of OR2 signal becomes larger than the OR1 signal from the point where the OR3 signal begins to appear. Similar to what is seen in EIT studies, the linewidth---of order few MHz---is determined by the decoherence and interference of the dressed states, and {\em not the decoherence of the lower levels} which are taken to have a fixed decay rate of $2 \pi \times 0.1$ MHz.

 {In order to see the importance of the differential dressing of the $\Lambda$-type and V-type systems}, we have repeated the calculation without including the $\Lambda$-type system. This is easy to implement in the calculation because it just requires setting $\Omega_{13}$ in the density-matrix equations to be 0. These results are shown in the insets of Fig.\ \ref{theor}. As seen, only the primary rotation appears, while both OR2 and OR3 do not appear even when the Rabi frequency is increased to $2.1 \, \Gamma$,  {thus highlighting the significance of the double dressing.}

But there are two problematic features in the theoretical results: (i) the calculated rotations appear at a lower value of $\Omega_c$, and (ii) they have a linewidth that is significantly smaller than the observed one. The explanation for (i) is that first it is difficult to relate the measured power to what is actually seen by the atoms because of absorption both at the cell walls and by preceding atoms; and second the spatial intensity profile of the beam is Gaussian and the Rabi frequency varies across the beam, while the model just assumes a single intensity. The explanation for (ii) is that first the observed linewidth is given by the {\em convolution} of the actual linewidth with the linewidth of the probe laser; and second the calculation does not account for noise that is always present in an experiment. Thus, until the actual linewidth is more than about $0.5$ MHz, experimental factors of finite probe linewidth and finite signal-to-noise ratio will prevent it from being observed.

\section{Conclusion}
In summary, we have studied polarization rotation of a linearly polarized probe beam passing through Rb vapor {\em in the absence of any magnetic field}. The transition used is the $F=2\rightarrow F'=3$ hyperfine transition in the $D_2$ line of $^{87}$Rb. The rotation is induced by a strong circularly polarized control laser driving the same transition, with the quantization axis set by its direction of propagation. The control laser plays a similar role as that in EIT and EIA experiments in the sense that it is used to modify the properties of the medium for the probe laser, but the difference is that EIT and EIA experiments depend on control-induced modification to the absorption in the medium (imaginary part of the susceptibility), whereas rotation experiments depend on control-induced modification to the dispersion in the medium (real part of the susceptibility). The lineshape of the rotation resonance is therefore dispersive.

The experimental spectra show a primary rotation signal at low control powers. Two higher-order rotations appear nested within this signal as the power is increased. The third-order rotation signal starts appearing when the control Rabi frequency increases beyond $1.5 \, \Gamma$, and has a subnatural linewidth as small as $0.14 \, \Gamma$. The explanation is that, due to optical pumping among the ground-state magnetic sublevels by the control laser, all the population is in the extreme sublevel. The complicated sublevel structure is then simplified to a $\Lambda$V-type system.  {The higher-order rotation signals arise because of the differential dressing of the sublevels coupled by the control laser, arising due to the different transition strengths. This explanation is borne out by a density-matrix analysis of the $\Lambda$V-type system, with and without including the $\Lambda$-type part.} The dispersive lineshape and subnatural linewidth make the higher-order resonances suitable for a variety of applications in precision measurements.

\begin{acknowledgments}
This work was supportedby the Department of Science and Technology, India.
\end{acknowledgments}

%\bibliographystyle{eplbib}
%\bibliographystyle{osajnl}
%\bibliography{D:/papers/eitsubnat/eitrefs}
%\end{document}


\end{document}